\documentclass[12pt]{article}
\usepackage{amssymb}
\usepackage{amsmath}
\usepackage{epsfig}
\usepackage{float}
\newcommand{\be}{\begin{equation}}
\newcommand{\ee}{\end{equation}}
\newcommand{\bea}{\begin{eqnarray}}
\newcommand{\eea}{\end{eqnarray}}
\begin{document}

\begin{center}
{\bf BRUNO PONTECORVO: MISTER NEUTRINO} \footnote{A report at 
III International Workshop 
"NEUTRINO OSCILLATIONS IN VENICE", Venice, February 7-10, 2006.}

\end{center}

\begin{center}
S. M. Bilenky
\end{center}

\begin{center}
{\em  Joint Institute
for Nuclear Research, Dubna, R-141980, Russia, and\\
SISSA,via Beirut 2-4, I-34014 Trieste, Italy.}
\end{center}

Bruno Pontecorvo was a great physicist who had enormous impact on the development of neutrino 
physics.
He was charming, marvelous person. I worked with B. Pontecorvo in Dubna 
about 15 years and during these years we were connected by warm, friendly relations. 
In  this talk I will speak about 
major contributions of Bruno Pontecorvo to neutrino physics and I will try to 
convey my impression about  his personality.
\section{Radiochemical method of neutrino detection}
In 1933-34 after a famous Pauli idea of neutrino and the discovery of the neutron E. 
Fermi proposed the 
first theory of the 
$\beta$- decay of nuclei. Fermi assumed that the Hamiltonian of the process 
\be
n\to p +e^-
+\bar \nu_{e}
\label{1}
\ee
has the vector form
\be 
\mathcal{H}_{\beta}= G_{F}\bar p\gamma_{\alpha}n ~\bar e  \gamma^{\alpha}  \nu_{e}       +\rm{h.c.}  
\label{2}
\ee
After Fermi's work in 1934 Bethe and Peierls made the first calculation of the cross section of interaction of neutrino with nuclei. They connected the cross section of the process 
$\bar\nu_{e}+(A,Z+1) \to e^{+}+ (A,Z)$
with the life-time  of the corresponding $\beta$-decay 
$(A,Z) \to e^{-} +\bar\nu_{e} + (A,Z+1)$
and showed that the cross section is extremely small ( $\simeq 10^{-43} \rm{cm}^{2}$ at MeV energies). There was no methods at that time which would allow to measure such small cross sections. During more than 10 years {\em neutrino was considered as undetectable particle}.
W.Pauli wrote to his friend ``....I have predicted something which shall never be detected experimentally''

{\em The first method of neutrino detection was proposed by B. Pontecorvo in 1946}  (Chalk River Laboratory report ``Inverse $\beta$ process'' \cite{BP1}).  In this paper he wrote

``It has been currently stated in the literature that inverse $\beta$-processes produced by neutrinos can not be observed, due to the low yield.'' 

And further

``The object of this note is to show that experimental observation of neutrinos is not out of question and to suggest a method which might make an experimental observation feasible''

B. Pontecorvo considered the process

$$\rm{``neutrino''} + (A,Z) \to e^- +(A,Z+1) $$

At that time it was not known whether neutrino and antineutrino are the same or different particles.
As a possible source of  neutrinos  B. Pontecorvo considered reactor, the sun and radioactive materials which can be extracted from reactor. The main idea of the proposed method was formulated as follows

``radioactivity of the produced nucleus may be looked for as a proof of the inverse $\beta$ process''.

B. Pontecorvo considered several reactions which can be used for neutrino detection. The 
process 
\be
\nu_{e}+ ^{37}\rm{Cl} \to e^- +^{37}\rm{Ar}, 
\label{3}
\ee
which later was named  Pontecorvo-Davis reaction, he considered as the most promising one. 
 He wrote

``The experiment with Chlorine, for example, would consist in irradiating with neutrinos a large volume of Carbon Tetra-Chloride for the time of the order of one month and extracting the radioactive  
$^{37}\rm{Ar}$ from the volume. The radioactive argon would be introduced into a small counter, 
the counting efficiency is close to 100\%, because of the high Auger electron yield.''

In 1948 B.Pontecorvo discovered the new high gas gain (up to $10^{6}$) proportional regime and proposed low-background 
proportional counter with high amplification \cite{BP2}

Pontecorvo $^{37}\rm{Cl} -^{37}\rm{Ar}$ method was realized by  R. Davis in  his pioneer 
solar neutrino experiment for which he was awarded in 2002  the  Nobel Prize.
Pontecorvo radiochemical method was used in the GALLEX and SAGE experiments.
The Pontecorvo counter was
crucial for detection of neutrinos in Homestake, GALLEX and SAGE experiments. 
\section{Universal weak interaction}
After the famous Conversi, Pancini, Picconi experiment  \cite{Conversi} in which it was discovered that muon is not 
strongly interacting Yukawa particle, B.  Pontecorvo became fascinated by the muon. In the paper 
\cite{}  he noticed that the probabilities of $\mu$-capture by nuclei and electron K-capture 
are practically the same  (if kinematics and initial wave function effects are taken into account).
He came to the conclusion that in muon capture neutrino is emitted and the spin of muon is equal 
1/2. B. Pontecorvo concluded in \cite{BP3} that 
``there exists
fundamental analogy between $\beta$-
processes and processes of emission and absorption of charged mesons'' (muons) 
He was the first who came to the idea 
of the existence of an universal weak interaction which includes 
interaction of nucleons with $e-\nu$ and $\mu-\nu$ pairs. 
Later the hypothesis of $\mu-e$ universality of weak interaction was proposed by other authors 
\cite{Puppi}.

In 1947-49 in Canada B. Pontecorvo and E. Hincks made  pioneer experiments on the investigation of muon decay \cite{BP4,BP5}.
They   found
\begin{enumerate}

\item that the charge particle produced in the muon decay is electron 
(measuring the intensity of the bremsstrahlung radiation)

\item that the decay $\mu \to e +\gamma$ is forbidden
(searching for electron-photon coincidence)

\item that the muon decays into three particles (measuring electron spectrum).

\end{enumerate}

\section{ Accelerator neutrinos, $\nu_{\mu} \not=\nu_{e}$}
In the end of the fifties there was a plan in Dubna to build a meson factory. In 1959 B. Pontecorvo 
started to think about the possibilities to do neutrino experiments at meson factories and high energy accelerators. He came to an idea of the feasibility of
experiments with accelerator neutrinos. He understood that one of the first problem 
which can be solved in such experiments is the problem of existence of two types of neutrinos 
(electron and muon). 

From the time of the investigation of the muon decay B. Pontecorvo 
knew that neutrinos which are produced in the decay of muon could be different
(at that time they even have different names: neutrino and neutretto, $\nu$
and $\nu'$ etc). In  the end of the fifties there was a  model dependent indication, 
coming from the limit on $\mu \to e \gamma$, that $\nu_{\mu} $ and $\nu_{e}$
are different \cite{Feinberg}.

In 1959 B.P. proposed an accelerator experiment which could allow to obtain
the decisive answer on question of the existence of the  second 
type of neutrino \cite{BP6}.
His proposal was realized in the famous Brookhaven experiment in 1962 \cite{Brookhaven}.
In 1988 L.Lederman, J. Steinberger and M. Schwartz were awarded the Nobel prize for the discovery of the 
muon type of neutrino.

\section{The birth of neutrino oscillations}

{\em B. Pontecorvo was  the pioneer of neutrino oscillations}. He came to the idea of neutrino oscillations in 
1957 soon after the two-component neutrino theory was proposed by Landau, Lee and Yang and Salam
\cite{Landau}  and confirmed by  Goldhaber et al  experiment \cite{Goldhaber}. The two-component neutrino theory is based on the assumption that neutrino is a massless particle and at that time (and many years later) massless neutrino was a prevailing idea of many physicists.

B. Pontecorvo believed in the analogy between  the weak interaction of hadrons (quarks) and leptons
and he believed that in the lepton world there exists phenomenon similar to $K^{0}\leftrightarrows     \bar K^{0}   $ 
oscillations. For the first time he  mentioned the possibility of neutrino oscillations 
in the paper 
\cite{BP7} in which he considered transitions $\mu^{+}e^{-}\to \mu^{-}e^{+}$ :

``If the two-component neutrino theory turn out to be incorrect
and if the conservation law of neutrino charge would not apply,
then in principle neutrino $\leftrightarrows $ antineutrino transitions
could take place in vacuum.''

The first paper on neutrino oscillations was published by Bruno Pontecorvo 
in 1957\cite{BP8} At that time F. Reines and C. Cowan \cite{Reines} successfully discovered antineutrino via 
the observation of the process
\be
\bar \nu_{e} +p \to e^{+} +n
\label{4}
\ee
At the same time R.Davis \cite{Davis} was doing the experiment with reactor antineutrinos searching for the process
\be
\bar \nu_{e} +^{37}\rm{Cl} \to e^{-} +^{37}\rm{Ar}
\label{5}
\ee
A rumor reached  B. Pontecorvo that Davis observed events (\ref{5}). He published  
the paper \cite{BP8} dedicated to neutrino oscillations.  He made the following two basic assumptions:

\begin{enumerate}

\item Neutrino and antineutrino emitted in $\beta$-processes are different particles.

\item Exists some interaction  which do not  conserve lepton number.
\end {enumerate}
He wrote in \cite{BP8}:
``It follows from 1. and 2. that neutrinos in vacuum can transform themselves into antineutrinos and vice versa.  This means that neutrino and antineutrino are particle mixtures , i.e., a symmetric and
antisymmetric combination of two truly neutral Majorana particles $\nu_1$
and $\nu_2$. ``
And further in \cite{BP8}:
`...`a beam of neutral leptons from a reactor which at first consists mainly of antineutrinos will change its composition and at certain distance $R$
from the reactor will be composed of neutrino and antineutrino in equal quantities....; Thus,
the cross section of the
production of neutrons and positrons in the process of the absorption of antineutrinos from a reactor by protons (experiment of Reines and Cowan) must  be smaller than the expected cross section.''

B. Pontecorvo strongly believed in neutrino oscillations.
From 1957 
he became
great enthusiast of neutrino oscillations.  For the rest of his life
neutrino oscillations were his beloved subject.

From 1958 paper

``effects of transformation of neutrinos into antineutrinos and vice versa may be unobservable
in the laboratory because of large values of $R$ but will certainly occur, at least on an astronomical scale''

I would like to stress that it was very nontrivial to propose neutrino oscillations in 1957,  
 at the time when only one type of neutrino  was known. Oscillations,  which B. Pontecorvo
 considered,  were $\nu_{L} \rightleftarrows \bar \nu_{L} $ i.e. oscillations 
between active and sterile neutrinos. He proposed not only  neutrino oscillations but 
 also existence of sterile 
neutrinos. By the way, the terminology``sterile neutrino'', so popular nowadays, 
 was invented by  B. Pontecorvo in 1967.

In the sixties  B. Pontecorvo discussed the problem of neutrino mass with L. Landau,  one of 
the authors of the theory of two-component massless neutrino \cite{BP9}. 
After Feynman and Gell-Mann \cite{Feynman}, Marshak and Sudarshan \cite{Marshak}  V-A theory Landau changed 
his opinion about neutrino mass. From his point of view after V-A theory it was  very natural 
to assume that neutrinos, like other fermions, have  nonzero masses. He 
supported Pontecorvo idea of small neutrino masses and oscillations.

After discovery of $\nu_{\mu}$ 
it was natural (and not difficult) for 
B. Pontecorvo to generalize his idea of neutrino oscillations
for the case of two types of neutrinos. In 1967 paper \cite{BP10}
he considered all possible transitions between two types of neutrinos: 
$\nu_{\mu} \rightleftarrows \nu_{e}$,
transitions into sterile neutrinos $\nu_{\mu} \leftrightarrows \bar \nu_{\mu L}$ etc.

Before R. Davis published his first result on the detection of solar neutrinos 
B. Pontecorvo wrote \cite{BP10}

``From observational point of view the ideal object is sun. If the oscillation length is smaller than the radius of the sun region effectively producing neutrinos
direct oscillations will be smeared out and unobservable. The only effect on the earth's 
suffice would be that the flux of observable solar neutrinos must 
be two times smaller than the total neutrino flux''.

{\em B. Pontecorvo envisaged the solar neutrino problem}:

``the question of lepton conservation has a bearing on the interpretation of the first experiments which will be soon performed with sun neutrinos''

The paper of V. Gribov and B. Pontecorvo \cite{Gribov} was an 
 important step in the development of the theory of neutrino oscillations. It was an opinion at that time that with left-handed neutrino fields 
 $\nu_{eL}$ and$\nu_{\mu L}$ it is 
impossible to  introduce neutrino masses. It is correct if total lepton number is conserved. 
 V. Gribov and B. Pontecorvo showed that if total lepton number is violated it is possible to 
introduce neutrino (Majorana) mass term. In such a scheme it will be transitions only between 
active neutrinos and antineutrinos. They applied the developed formalism of two-neutrino oscillations to solar neutrinos.

In retrospect  B. Pontecorvo had a chain of connected neutrino 
ideas which he developed all his scientific life. In 1946 he proposed radiochemical method of neutrino detection which allowed 
R.Davis in his pioneer experiment to discover solar neutrinos. He was the first who understood that 
reactor and the sun are sources of neutrinos which could be detected in an experiment. 
In 1947 he came to an idea of the existence of $\mu-e$ universal weak inteaction. He was one of the first who understood the feasibility of 
experiments with accelerator neutrino and he proposed the experiment which
allowed to proof that $\nu_{\mu}$ and $\nu_{e}$ are different particles. 
In 1958 he came to the idea of neutrino oscillations.

\section{Bruno Pontecorvo. Some recollections}
B. Pontecorvo liked very much underwater fishing. 
Usually he catch-ed fish (may be a lot)  in summer in Crimea during vacation.
An Italian friend gave him as a present the Calipso suit. So he could enjoy 
underwater fishing also in autumn in small rivers near Moscow.
He liked the river Nerl about two hours driving  from Dubna.
Usually he invited my wife and me for such trips. Sometimes his wife Marian also was with us.
While he was in a river
we made fire and collected mushrooms.
Often  B.P. went out from  river without fish...
fire and mushrooms were there.
Only once during very dry summer, when fires were forbidden, Bruno captured a lot of fish.

We started our collaboration on neutrino oscillations in 1975  in a car during such trip.
We published many papers and conference reports. In 1977 we wrote 
the first review on neutrino oscillations\cite{BP} which attracted attention of many physicists to the problem 
of neutrino masses and oscillations.
Our last review was written in 1987 
 for the Italian Encyclopedia ``Scienza e Tecnica''.

We considered all possible neutrino mass terms (Dirac, Majorana, Dirac and Majorana) and different phenomenological possibilities to reveal effects of small neutrino masses.
 We did not try to build the theory of neutrino masses.
\footnote{ After long discussions in one of our first paper \cite{BilP} we concluded, however.
``It seems to us that the special cases of mixing angle $\theta=0$ and 
$\theta=\pi/4$ are of the greatest interest''. This expectation is not very far from the present-day situation with large $\theta_{12}$ and $\theta_{23}$ and small $\theta_{13}$}
Our general point of view was the following
\begin{itemize}
\item
We do not know the values of neutrino masses. Neutrino oscillations as an interference 
phenomenon are sensitive to extremely small neutrino mass-squared differences. 
Investigation of neutrino oscillations give us an unique possibility 
to study the problem of small neutrino masses and neutrino mixing.
\item
Neutrino experiments with neutrinos from different facilities are sensitive to 
different values of neutrino mass-squared differences. Neutrino oscillations must be search for in all possible neutrino experiments.
\end{itemize}
This strategy brought success. 
It required many years and heroic efforts of many experimental groups \cite{neutrinos} to reveal 
effects of tiny neutrino masses.
{\em The discovery of neutrino oscillations 
 was real triumph of Bruno  Pontecorvo}  who proposed neutrino oscillations and pursued 
the idea of oscillations 
for many years, when the general opinion favored massless neutrinos and no neutrino oscillations.

The years of work and friendship with Bruno Pontecorvo were the happiest and unforgettable
years in my life.
His wide and profound knowledge of physics, his love of physics, his ingenious intuition and
his ability to understand complicated problems in a clear and simple way
were gifts of God.
Bruno Pontecorvo was a true scientist in the best, classical sense of the word.
When he thought about some problem he thought about it continuously from early  morning till late evening. He devoted all his resources and great intellect  to science, and 
though he was not indifferent to  the recognition of his contribution to physics, his main stimulus was 
the search for the truth

More than ten last years were for Bruno Pontecorvo  years of  courages struggle against 
Parkinson disease. His love to physics and to neutrino helped him to overcome 
difficult problems of the illness. He  never stopped to work, to think about neutrinos and 
to continue active life.

 Two days before his death he  was in his office at the second floor of the Laboratory of Nuclear Problem in Dubna, where he spent 43 years.
When he was leaving the Laboratory for the last time he looked through the  window 
and said to his secretary Irina Pokrovskaja: ``Look how beautiful are these colours....''
It was nice Russian September 1993.

Bruno Pontecorvo was born in Pisa in 1913. He entered the Engineer Faculty of the Pisa University. 
After two years he decided to study physics and joined faculty of Physics and Mathematics of Rome University. From 1931 till 1936 first as a student and then as a researcher he worked in the famous Fermi group and was one of ``ragazzi da via Pansperna''.  B. Pontecorvo participated in the discovery of the effect of slow 
neutrons. In 1936-40 B. Pontecorvo worked on nuclear isomerism in Joliot-Curie group  
in Paris. In 1940-42 he worked in USA. He developed the neutron well-logging, an effective method of 
prospering for oil. In 1943-48 he worked in Canada. He took part in design and 
starting up of a heavy water uranium research reactor. In Canada B. Pontecorvo made muon experiments and started to work on neutrinos. In 1948-50 he worked in Harwell and from September 1950 he worked in Dubna.

Bruno had four  brothers and 
three sisters. One of his brother Guido was famous biologist. Gillo Pontecorvo is famous  
film director. Opinion of parents.... 
``Guido era il piu intelligente dei fratelli, Paolo era il piu' serio,
Giuliana la piu  colta, Bruno il piu' buono ma il piu' limitato, come era dimonstrato dai suoi occhi buoni ma non intelligenti...'' (B. Pontecorvo. Una nota autobiografica)
``Guido was the most intelligent among the brothers, Paolo was the most serious, Giuliana was the 
most cultured. 
Bruno was the most gentle but the most limited , as one could understand from his eyes sweet but not intelligent...''

Bruno Pontecorvo was very friendly, nice, highly intelligent person. Everybody who was familiar with him love him. He had many friends in Italy, Russia and other countries. He liked cinema, theater, music, literature, tennis, football,...In 1988 when he was 75 there was 
a big celebration in Dubna. Many of Bruno friends came. There were many talks and many jokes. Some 
pictures from that celebration drawn  by Misha Bilenky can be seen below.

I acknowledge the support of  the Italian Program  ``Rientro dei cervelli''.
It is my pleasure to thank I. Todorov for careful reading of the paper and numerous remarks.

\begin{figure}[!t]
\centerline{\epsfig{file=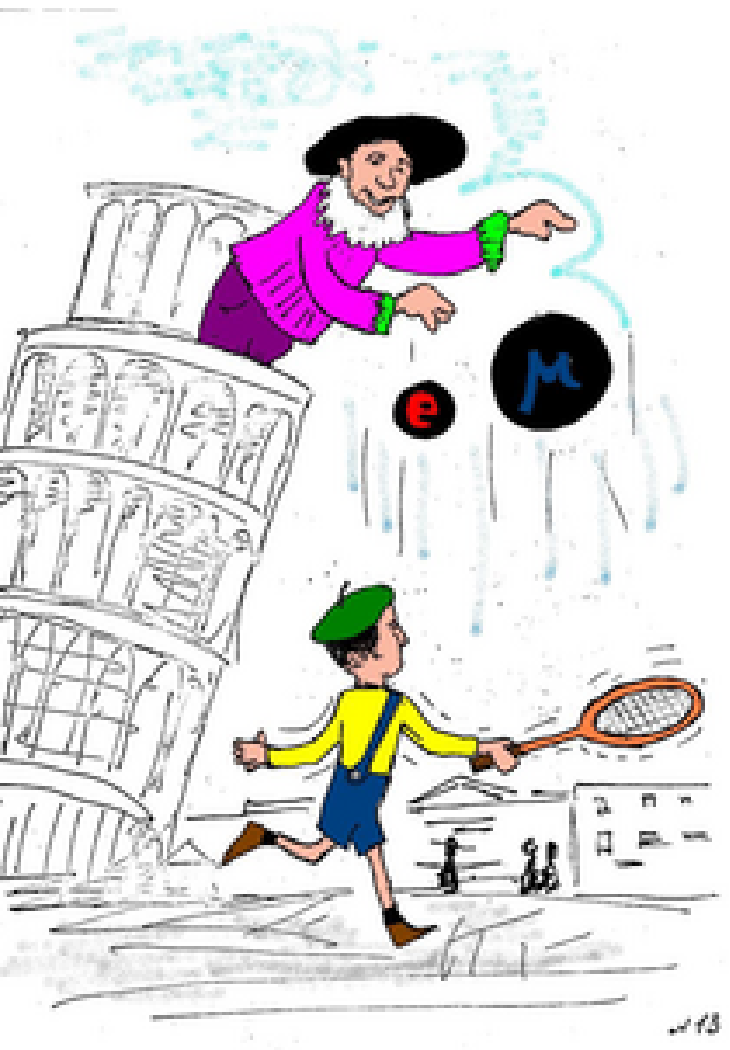,width=0.9\linewidth}}
\caption{``A scuola ero bravo ma la cosa piu importante nella mia vita era il tennis'' 
(B. Pontecorvo. Una nota autobiografica)
\label{fig.1}}
\end{figure}

\begin{figure}[!t]
\centerline{\epsfig{file=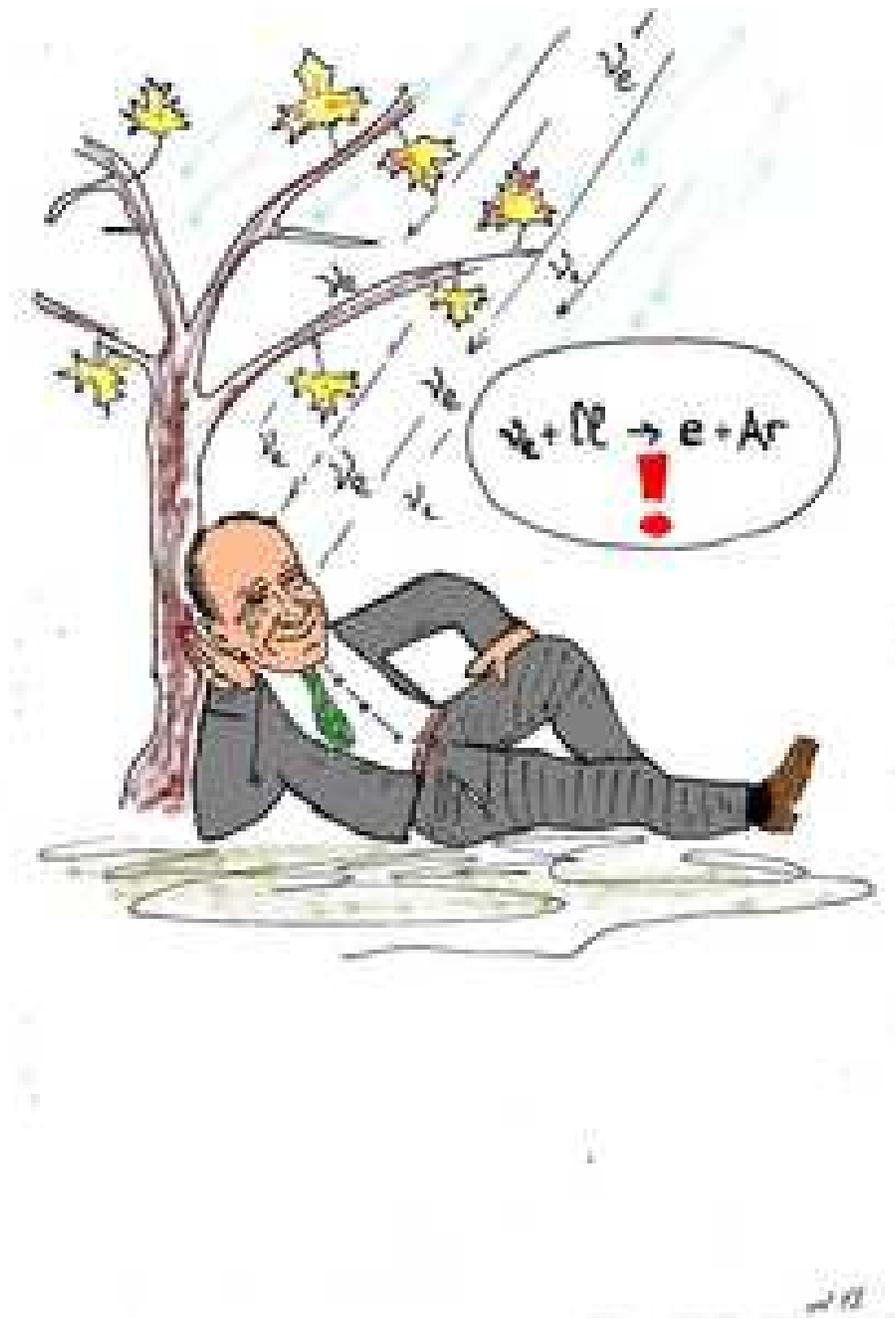,width=0.9\linewidth}}
\caption{In Canada.Chlorine-Argon idea. (Remember the story about Newton and apple).
\label{fig.1}}
\end{figure}

\begin{figure}[!t]
\centerline{\epsfig{file=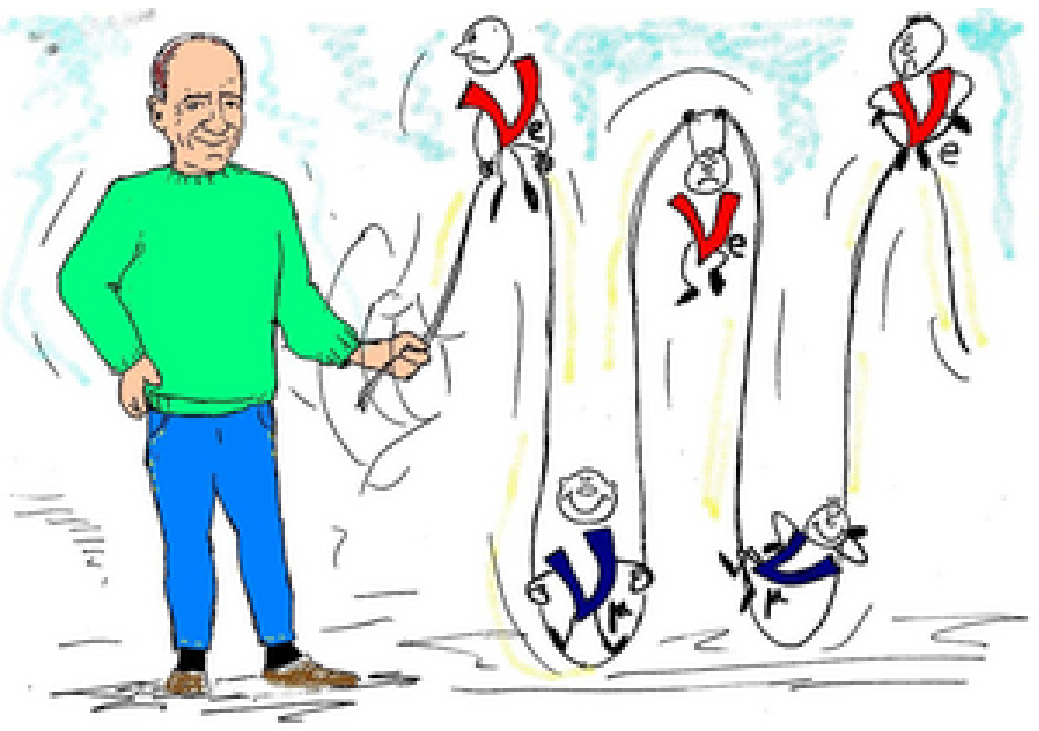,width=0.9\linewidth}}
\caption{ Dubna 1988. Neutrino oscillations.
\label{fig.1}}
\end{figure}


\begin{thebibliography}{99}

\bibitem{BP1} B. Pontecorvo,  Report PD-205, Chalk River Laboratory, 1946.
see B. Pontecorvo ``Selected scientific works''   p.21, Societa Italiana di Fisica, Bologna, Italia. 

\bibitem{BP2} B. Pontecorvo, Helv. Phys. Acta Suppl.  {\bf 3} (1950) 97.


\bibitem{Conversi} M. Conversi , E. Pancini and G. Piccioni Phys. Rev.   {\bf 71} (1947) 209.





\bibitem{BP3} B. Pontecorvo, Phys. Rev.   {\bf 72} (1947) 246.



\bibitem{Puppi} O. Klein, Nature  {\bf 161} (1948) 897; G. Puppi, Nuovo Cimento 
{\bf 5} (1948) 587




\bibitem{BP4} E. Hinks and B. Pontecorvo, Phys. Rev.   {\bf 73} (1948) 257; 
Can. J. Res. {\bf 28A} (1950) 29.


\bibitem{BP5} E. Hinks and B. Pontecorvo, Phys. Rev.   {\bf 75} (1949) 268; 
Phys. Rev.   {\bf 77} (1950) 102 



\bibitem{BP6}B. Pontecorvo, 
J.Exptl. Theoret. Phys. {\bf 37} (1959) 1751.



\bibitem{Feinberg} G. Feinberg, Phys. Rev.   {\bf 110} (1958) 1482. 



\bibitem{Brookhaven} G. Danby, J.M. Gaillard, K. Goulianos, L.M. Lederman, M.
Mistry, M. Schwartz and J. Steinberger, Phys. Rev. Lett.
{\bf 9} (1962) 36.


\bibitem{Landau} L.D. Landau, Nucl.Phys.{\bf 3} (1957) 127; T.D. Lee and C.N.Yang, 
Phys. Rev. {\bf 105} (1957) 1671; A. Salam, Nuovo Cimento {\bf 5} (1957) 299.


\bibitem{Goldhaber} M. Goldhaber, L. Grodzins and A.W. Sunyar, Phys.
Rev. {\bf 109} (1958) 1015 .



\bibitem{BP7}B. Pontecorvo, 
J.Exptl. Theoret. Phys. {\bf 33} (1957) 549.


\bibitem{BP8}B. Pontecorvo, J.Exptl. Theoret.
Phys. {\bf 34} (1958) 247 [Sov. Phys. JETP {\bf 7} (1958) 172].






\bibitem{Reines} 
F. Reines and C. Cowan, Nature  {\bf 178} (1956) 446.; C. Cowan, F. Reines et. al , Science  
{\bf 124} (1956) 103 ;
F. Reines and C. Cowan,  Phys.Rev. {\bf 113} (1959) 273; F. Reines et al., Phys. Rev.{\bf  117} (1960) 159 . 

\bibitem{Davis} R. Davis, {\em Bull. Am. Phys. Soc.} ( Washington.
meeting, 1959).


\bibitem{BP9} B.Pontecorvo, private communication.


\bibitem{Feynman} R.P.Feynman and M.Gell-Mann, Phys.
Rev.{\bf 109} (1958) 193.


\bibitem{Marshak} E.C.G. Sudarshan and R. Marshak, Phys. Rev.
{\bf 109} (1958) 1860.


\bibitem{BP10} B.Pontecorvo, J.Exptl. Theoret.
Phys. {\bf 53} (1967) 1717.

\bibitem{Gribov} V. Gribov and B. Pontecorvo, Phys. Lett. {\bf B28} (1969) 493.


\bibitem{BP} S.M. Bilenky and B. Pontecorvo, Phys.
Rep. {\bf 41} (1978) 225 .


\bibitem{BilP}S.M. Bilenky and B. Pontecorvo,
 Phys. Lett. {\bf B61} (1976) 248; Yad.
Fiz. {\bf 3} (1976) 603 .

\bibitem{neutrinos} Homestake Collaboration, T. Cleveland {\it et al.}, Astrophys. J. {\bf
496} (1998) 505; Super-Kamiokande Collaboration, S.~Fukuda {\it et al.}, Phys. Rev. Lett.
 {\bf 86} (2001) 5651; GALLEX Collaboration, W. Hampel
{\it et al.}, Phys. Lett. {\bf B 447} (1999) 127 ;\,
GNO Collaboration,
M. Altmann {\it et al.}, Phys. Lett. {\bf B 490} (2000) 16 ;\,
Nucl.Phys.Proc.Suppl. {\bf 91} (2001) 44;
SAGE Collaboration,
J. N. Abdurashitov {\it et al.},
 Phys. Rev. {\bf C 60} (1999) 055801 ; \,Nucl.Phys.Proc.Suppl. {\bf 110}
(2002) 315; Super-Kamiokande Collaboration, Y. Ashie {\em et al.,} hep-ex/0501064;
Y.Fukuda {\em et al.,}  Phys.Rev.Lett.\textbf{87}(1998)1562;
Y. Ashie et al.,Phys. Rev. Lett. 93 (2004) 101801; SNO Collaboration, Phys.Rev.Lett.
\textbf{81} (2001) 071301;~\textbf{89} (2002) 011301
;~\textbf{89} (2002) 011302;~
nucl-ex/0502021;KamLAND Collaboration,
T.Araki {\em et al.},
Phys.Rev.Lett. \textbf{94} (2005) 081801;~
hep-ex/0406035; K2K Collaboration, E. Aliu {\em et al.},
Phys.Rev.Lett. \textbf{94}(2005) 081802.


































\end{thebibliography}
 \end{document}